\documentclass{article}
\usepackage[margin=1in]{geometry}

\usepackage[T1]{fontenc}
\usepackage[utf8]{inputenc}
\usepackage{cite}
\usepackage{url}
\usepackage{amsmath,amssymb,amsfonts,amsthm}
\usepackage{mathptmx}
\usepackage{color}
\usepackage{float}
\usepackage{tikz}
\usepackage{float}
\usepackage{readarray}
\usepackage{fontawesome5}
\usepackage{algorithm,algorithmic}
\usepackage{mathtools}
\usepackage{graphicx}
\usepackage{textcomp}
\usepackage{xspace}
\usepackage[pdfusetitle]{hyperref}

\usepackage{graphicx}
\usepackage[inkscapelatex=false]{svg} \usepackage{tikz}
\usetikzlibrary{automata, positioning, arrows, calc}
\usepackage{caption}

\usepackage{etoolbox}

\BeforeBeginEnvironment{enumerate}{\begin{samepage}}
\AfterEndEnvironment{enumerate}{\end{samepage}}

\theoremstyle{definition}
\newtheorem{definition}{Definition}[section]

\theoremstyle{remark}
\newtheorem{remark}{Remark}[section]
 \tikzset{
  gadget/.style={
    ->, 
    >=stealth',
    auto, 
    line width=1.2pt, 
    node distance=3cm,
    every state/.style={
      draw=black, 
      line width=1.2pt
    }
  }
}

\makeatletter
\def\fps@figure{htbp}
\makeatother

\newcommand{\changedef}[1]{\ifmmode
    \begingroup\color{blue}#1\endgroup \else
    \textcolor{blue}{#1}\fi
}

\newcommand{\sysname}{\textsc{Gadgeteer}\xspace}

\newcommand{\locfmt}[1]{$\ell_{\text{#1}}$}
\newcommand{\statefmt}[1]{$q_{\text{#1}}$}
 
\title{Push-1 is PSPACE-complete, and the automated verification of motion planning gadgets}
\author{
Zachary DeStefano\thanks{\href{mailto:zd@nyu.edu}{zd@nyu.edu}}\and
Bufang Liang\thanks{\href{mailto:bufangliang@gmail.com}{bufangliang@gmail.com}}}
\date{}

\begin{document}

\maketitle

\begin{abstract}
Push-1 is one of the simplest abstract frameworks for motion planning; however,
the complexity of deciding if a Push-1 problem can be solved was a several-decade-old open question.
We resolve the complexity of the motion planning problem Push-1 by showing that it is PSPACE-complete,
and we formally verify the correctness of our constructions.
Our results build upon a recent work which demonstrated that Push-1F (a variant of Push-1 with fixed blocks) and Push-k for $k \geq 2$ (a variant of Push-1 where the agent can push $k$ blocks at once) are PSPACE-complete and more generally on the motion-planning-though-gadgets framework.

In the process of resolving this open problem, we make two general contributions to the motion planning complexity theory.
First, our proof technique extends the standard motion planning framework by assigning the agent a state.
This state is preserved when traversing between gadgets but can change when taking transitions in gadgets.
Second, we designed and implemented a system, \sysname, for computationally verifying the behavior of systems of gadgets.
This system is agnostic to the underlying motion planning problem, and allows for formally verifying the correspondence between a low-level construction and a high-level system of gadgets as well as automatically synthesizing gadgets from low-level constructions.
In the case of Push-1, we use this system to formally prove that our constructions match our high-level specifications of their behavior.
This culminates in the construction and verification of a self-closing door, as deciding reachability in a system of self-closing doors is PSPACE-complete.
\end{abstract}
 \section{Introduction}

Over the past few decades, a great deal of research has gone into analyzing the complexity of various abstract block-pushing related motion planning problems on a unit-square grid, colloquially referred to as the \emph{Push} family of problems.
Broadly speaking, these problems have a designated goal square, obstacles (called blocks), and an agent.
The agent has the task of reaching the goal square, can move into adjacent unoccupied (empty) grid squares, and can push blocks, with some restrictions, (including pushing blocks over top of the goal square).

\begin{figure}
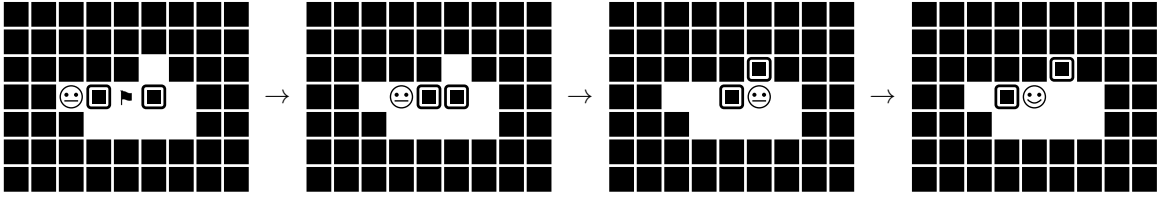

  \centering
  \begin{minipage}[c]{0.2\textwidth}
    \centering
    \includesvg[width=\linewidth]{svgs/level/level1.svg}
  \end{minipage}
  \;$\to$\;
  \begin{minipage}[c]{0.2\textwidth}
    \centering
    \includesvg[width=\linewidth]{svgs/level/level2.svg}
  \end{minipage}
  \;$\to$\;
  \begin{minipage}[c]{0.2\textwidth}
    \centering
    \includesvg[width=\linewidth]{svgs/level/level3.svg}
  \end{minipage}
  \;$\to$\;
  \begin{minipage}[c]{0.2\textwidth}
    \centering
    \includesvg[width=\linewidth]{svgs/level/level4.svg}
  \end{minipage}
  \caption{Example Push-1 level and shortest solution (shown as a series of pushes).
  The starting state is on the left, and each push is shown.
  For visual clarity, blocks in the start state which cannot be moved at any point in time, because they
  form a $2 \times 2$ grid, are depicted differently than blocks that can be pushed, even though they are the same type of object.
  Recall that the agent cannot push two or more adjacent blocks and that blocks can be pushed onto and off of the goal square.}
  \label{fig:level}
\end{figure}

Within the Push family, there are several variations of this basic ruleset, all modifying the restrictions on pushing blocks.
Push-$k$ allows the agent to push a row or column of at most $k$ blocks at once, as long as there is an empty grid square at the end of the sequence.
Push-\textasteriskcentered\ takes this to the extreme, allowing for any number of blocks to be pushed at once.
Adding the \emph{F} modifier, for example, Push-1F, allows for certain blocks to be designated as ``fixed'' or immovable.
Figure~\ref{fig:level} shows an example Push-1 problem and solution.

For a thorough overview of the state of the art (prior to our contribution), we recommend \textit{Pushing Blocks via Checkable Gadgets}~\cite{ani22pushing} and Erik Demaine's subsequent talk on the subject at JCDCGGG 2022~\cite{demaine22pushing}.
The analysis of the complexity of these problems involves drawing a correspondence between low-level Push constructions
and high-level state machines called gadgets.
Recent proofs proceed by showing that it is possible to design a low-level construction which corresponds to a special type of gadget, called a self-closing door, because deciding reachability in planar assemblies of these doors is PSPACE-complete~\cite{ani20walking}.

Prior to writing this paper, there were two remaining unresolved questions about the complexity of deciding reachability of two problem types in the Push family, Push-1 and Push-\textasteriskcentered.
These are both known to be NP-hard~\cite{demaine00push1, hoffmann00pushstar, demaine03pushstar}, and though it was conjectured that Push-1 is PSPACE-complete and Push-\textasteriskcentered\ is NP-complete, the exact complexities remained elusive.
\textit{Pushing Blocks via Checkable Gadgets}~\cite{ani22pushing} proved that
Push-2 and Push-1F were PSPACE-complete; however, the techniques did not seem to translate to proving the PSPACE-completeness of Push-1.

In this work, we resolve the complexity of Push-1, proving that it is PSPACE-complete, answering a 25-year-old question~\cite{demaine00pushpush, demaine00push1} and leaving the complexity of Push-\textasteriskcentered\ as the final open problem.
In the process of proving this result, we also make two more general contributions.
The first involves extending the classical motion-planning-through-gadgets framework~\cite{demaine18gadget}
allowing for the use of a wider variety of low-level constructions, some of which are crucial to the resulting proof.
The second involves designing and implementing a framework for computationally verifying the behavior of systems of gadgets.
Specifically, we mechanically verify that a complex combination of our low-level constructions exactly matches the high-level specification of a self-closing door.

In Section~\ref{s:gadget}, we provide background on the classical motion-planning-through-gadgets framework,
the extension to checkable gadgets proposed in \textit{Pushing Blocks via Checkable Gadgets}~\cite{ani20walking},
and our extension, which assigns the agent a state.
This state can be changed within gadgets and allows for information to be preserved and communicated between gadgets.
Somewhat counter-intuitively, this non-locality significantly simplifies the reasoning about systems of gadgets, and we believe it to be of independent interest in proving the complexity of other motion planning problems.
In the case of Push-1, this allows us to distinguish between a pure traversal between gadgets and a traversal in which the agent is pushing a block.
This non-local behavior, in combination with checkability, is central to our proof.

In Section~\ref{s:verif}, we provide details on the theory, mechanics, and implementation of our formal verification system.
First, we formalize what it means for a gadget and a construction to correspond,
and then we design a system to produce generalized proofs of correspondence between low-level constructions and high-level systems of gadgets for any type of motion planning problem.
To our knowledge, this is the first time anyone has proposed, designed, implemented, or applied such a tool for determining the complexity of motion planning problems.

In Section~\ref{s:proof}, we provide our proof of the PSPACE-completeness of Push-1.
This begins with the presentation of various low-level Push-1 constructions alongside their verified gadget descriptions.
It then progresses into combinations of these constructions and their verified combined behavior.
Finally, it culminates in the construction and verification of a self-closing door.
 \section{Gadget framework and our extension}
\label{s:gadget}

The motion-planning-through-gadgets framework
defined in \textit{Computational Complexity of Motion Planning of a Robot through Simple Gadgets}~\cite{demaine18gadget} and subsequently formalized and expanded by a variety of authors~\cite{lynch20framework, hendrickson21gizmo}
provides an abstraction for reasoning about and proving the hardness of motion planning problems.
Here we present the portions of the framework, specialized to a single-agent case, as relevant to this paper, followed by our contributions.
Central to this abstraction are gadgets that an agent can travel through and between.

\begin{definition}[Gadget, classical]
\label{def:gad1}
A \emph{gadget} $G$ is a 4-tuple $(Q, q_{\text{start}}, L, \Delta)$ where
$Q$ is a finite set of \emph{states},
$q_{\text{start}} \in Q$ is the initial state,
$L$ is a finite set of \emph{locations}, 
$\Delta \subseteq (Q \times L)^2$ is a set of directed \emph{transitions}.
These transitions, for example: $(q, \ell) \to (q', \ell')$
are interpreted as follows.
If the gadget is in state $q$ and an agent is at location $\ell$, then the agent can traverse to location $\ell'$, changing the state of the gadget to $q'$ in the process.
For visualization, it is convenient to visualize gadgets as finite state machines
and to write transitions in the form $q \xrightarrow{\ell \to \ell'} q'$.
\end{definition}

When an agent takes the transition $(q, \ell) \to (q', \ell')$, we may colloquially say
that the agent \emph{enters} at $\ell$ and \emph{exits} at $\ell'$.
Additionally, a transition is colloquially called trivial if it is of the form $(q, \ell) \to (q, \ell)$, that is, the agent enters and exits the same location without changing the state of the gadget.

\begin{figure}
  \centering
  \begin{minipage}[c]{0.35\textwidth}
    \centering
    \includesvg[width=\linewidth]{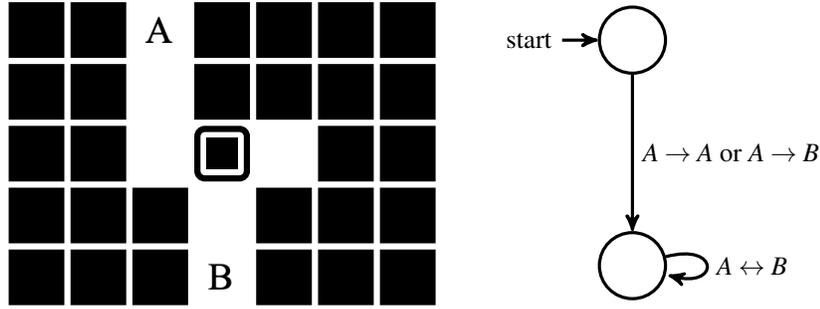}
  \end{minipage}
  \begin{minipage}[c]{0.35\textwidth}
    \centering
    \begin{tikzpicture}[gadget]
      \node[state, initial]    (0)                    {};
      \node[state]             (1) [below of=0]       {};
      \path (0) edge  node {$A \to A$ or $A \to B$} (1);
      \path (1) edge[loop right]  node {$A \leftrightarrow B$} (1);
    \end{tikzpicture}
  \end{minipage}
  \caption{Simple Push-1 construction (left) and gadget (right). All blocks in the construction, denoted by black squares, are movable; however, the only block that can be pushed at any point in time, by virtue of its position, is indicated with a special symbol. Trivial $A \to A$ and $B \to B$ transitions that do not change the state of the gadget are omitted for clarity. Additionally, transitions where the agent enters or exits with a block are also omitted for simplicity. The gadget shows that the $B \to A$ transition is only possible after the agent enters at $A$ and pushes the highlighted block one space to the right.}
  \label{fig:simp}
\end{figure}

Gadgets form a high-level abstraction and share a correspondence with low-level constructions, a notion we formalize in Section~\ref{s:verif}.
In the case of Push-1, Figure~\ref{fig:simp} depicts a low-level construction and its high-level gadget, side by side.

\begin{definition}[System of gadgets]
\label{def:sys}
A \emph{system of gadgets} is a pair $(H, E)$ where
$H$ is a set of gadgets,
and $E$ is a set of undirected edges between locations in the gadgets, called a \emph{connection graph}.
The edges in $E$ allow for an agent to traverse freely between locations on different gadgets.
\end{definition}

For the case of a planar system of gadgets,
the topology of a gadget is abstractly represented as a wheel graph of locations, and
there must exist an embedding of these gadgets and the connection graph in the plane without crossings.

\begin{definition}[Configuration]
\label{def:config}
A \emph{configuration} of a system of gadgets is the location of the agent and the respective states of all gadgets.
A \emph{valid transition} between configurations either consists of the agent taking a valid transition in a gadget or traversing an edge in the connection graph.
\end{definition}

In Section~\ref{s:verif}, we observe that it is possible to transform a system of gadgets into a single
gadget using some auxiliary information.
We believe this observation is implicit in prior works; however, we make it explicit in the process
of formal verification.

The reachability motion planning problem asks: given a system of gadgets, a fixed configuration, and a starting location $\ell_{\text{start}}$ for the agent, is there a sequence of valid transitions that results in a configuration where the agent is at location $\ell_{\text{final}}$?

Recent work has introduced \emph{checkable gadgets} in the problem of reachability.
It suffices here to provide the following simplified description of a checkable gadget,
which we refer to as a \emph{simply checkable gadget} to distinguish from the more general term.
A formal treatment of checkable gadgets can be found in \textit{Pushing Blocks via Checkable Gadgets}~\cite{ani22pushing}.

\begin{definition}[Simply checkable gadget]
A \emph{simply checkable gadget} is a gadget $G = (Q, q_{\text{start}}, L, \Delta)$ 
with a designated state $q_{\text{checked}}$, subset of states $\textbf{broken} \subset Q \setminus \{q_{\text{start}}, q_{\text{checked}}\}$, and locations $I, O \in L$ with the following two properties.
First, the only non-trivial transition in $G$ which exits at $O$ or leaves the state of the gadget as $q_{\text{checked}}$ is of the form $(q, I) \to (q_{\text{checked}}, O)$. These transitions are referred to as check-paths.
Second, there are no transitions of the form $(q, \ell) \to (q', \ell')$ where $q \in \textbf{broken}$ and $q' \not\in \textbf{broken}$.
Note: transitions where $q \not\in \textbf{broken}$ but  $q' \in \textbf{broken}$ are allowed.
\end{definition}

Using this definition, once the gadget is checked, it cannot be interacted with meaningfully later,
and once a gadget enters a \textbf{broken} state, the check-path is permanently unreachable.
The more general definition (omitted here) permits more complex check-paths consisting potentially of multiple transitions, all of which must be taken to check that the gadget has not entered a \textbf{broken} state.

If the reachability problem is designed such that all simply checkable gadgets must be in their respective $q_{\text{checked}}$ states,
for example, by requiring the agent to traverse a check-path in each
to reach $q_{\text{final}}$, then it is possible to ignore states in the \textbf{broken} set.
The process of eliminating/ignoring these states is called \emph{post-selection}. Several of the gadgets in Section~\ref{s:proof} require post-selection.

\paragraph{A modified gadget framework}

We modify these standard definitions by giving the agent
a state $s$ from a finite set of agent states $S$.
Additionally, it is helpful to designate a particular state, $s_{\text{start}}$, as the initial state of the agent.
Consider the following revised definition of a gadget with changes denoted in \changedef{blue}.

\begin{definition}[Gadget, new]
\label{def:gad2}
A \emph{gadget} $g$ is a 4-tuple $(Q, q_{\text{start}}, L, \Delta)$ where
$Q$ is a finite set of \emph{states},
$q_{\text{start}} \in Q$ is the initial state,
$L$ is a finite set of \emph{locations}, 
and $\Delta \subseteq (Q \times L \changedef{\times S})^2$ is a set of directed \emph{transitions}.
These transitions, for example: $(q, \ell\changedef{, s}) \rightarrow (q', \ell'\changedef{, s'})$
are interpreted as follows.
If the gadget is in state $q$ and an agent \changedef{in state $s$} is at location $\ell$, then the agent can traverse to location $\ell'$, changing \changedef{its own state to $s'$ and} the state of the gadget to $q'$ in the process.
For visualization, it is convenient to visualize gadgets as non-deterministic finite state machines
and to write transitions in the form $q \xrightarrow{\changedef{(}\ell\changedef{, s)} \to \changedef{(}\ell'\changedef{, s')}} q'$.
\end{definition}

Using this revised definition, a configuration of a system of gadgets is the location 
\changedef{and state} of the agent and the respective states of all gadgets.
When the set of agent states has size one, this revised definition trivially reduces to the 
classical framework.

We believe this extension more naturally captures real-world motion planning problems
and the full rulesets for setups described in prior works~\cite{demaine16mario, ani22pushing, demaine18portal, demaine18gadget, lily22celeste, hardness24mario} potentially allowing for simpler
proofs of hardness.
In many of these prior works, the agent has a state, but care is required to design gadgets
which either are not affected by this state or ensure that the agent returns to its original state at the end of each transition.
This extension helps to formalize reasoning about the correctness of this aspect of these gadgets.

\paragraph{Push-1, agent states, and topological restrictions}

The mechanics of Push-1 can be represented using the classical framework;
however, in this work, we distinguish between the agent stepping into an empty tile and an agent pushing a block.
This agent state ultimately allows for encoding the transfer of blocks between gadgets.
To avoid the possibility of the agent leaving a block in edges in the connection graph of a system of gadgets,
extra restrictions on the topology of a system must be enforced.
Specifically, if it is possible to a push a block between locations in two different gadgets, these locations must be directly adjacent in the square-grid, and the axis along which the block can exit one gadget must match the axis along which the block can enter the other.
If a block cannot be pushed between two gadgets, the edge connecting them has no such restrictions.

\paragraph{Self-closing doors}

The paper \textit{Walking through Doors is Hard}~\cite{ani20walking} provides a taxonomy of special gadgets called \emph{self-closing doors}.
For the purposes of this paper, it is useful to consider just one type of self-closing door
with $3$ locations (\locfmt{entrance}, \locfmt{exit}, and \locfmt{keyhole}), $2$ states (\statefmt{open} and \statefmt{closed}), and $2$ transitions.
If the door is in the \statefmt{open} state, the agent can travel from \locfmt{entrance} to \locfmt{exit}, changing the state of the door to \statefmt{closed} in the process. Hence the name ``self-closing''.
If the gadget is in the \statefmt{closed} state, the agent can start at and return to \locfmt{keyhole}, returning the door to the \statefmt{open} state.
The aforementioned paper proves that reachability in planar assembly of self-closing doors (of the form described above) is PSPACE-complete.
Thus, it suffices to construct a self-closing door in Push-1 to prove that reachability is PSPACE-complete.

 \section{Automated verification}
\label{s:verif}

The design and analysis of gadgets is complex and error-prone.
The descriptions given here and in Section~\ref{s:proof} may appear intuitive; however, very similar designs permit
radically different undesirable behaviors, often resulting from edge cases.
The presence of these easy-to-miss breaks, in particular in the development of the 1-toggle and precursor gadgets, are what motivated the computational verification.
This section describes the design and implementation of a system which automates this error-prone analysis.

\subsection{Formalizing correspondence}

Formalizing the correctness of the correspondence between a low-level construction
and a high-level gadget, requires first formalizing the definition of a low-level construction.
Consider the following definition.

\begin{definition}[Construction]
A \emph{construction} is a 5-tuple $(Q, q_{\text{start}}, L, P, \Delta)$ where
$Q$ is a finite set of states,
$q_{\text{start}} \in Q$ is the initial state,
$L$ is a finite set of locations, 
$P \subset L$ is a finite set of locations where the agent can enter and exit the construction, called \emph{ports}, and
$\Delta \subseteq (Q \times L \times S)^2$ is a set of directed transitions, where $S$ is the set of agent states.
\end{definition}

The transitions are typically not explicitly provided but instead are derivable from the rules of the construction.
For example, in Push-1, given a particular state and agent location, it is possible to derive the transition by looking at the contents of two consecutive grid-squares adjacent to the agent along each cardinal direction.
If the square adjacent to the agent is empty, it can move there.
If the square adjacent to the agent has a block but the subsequent square is empty, the agent can push the block.
Similarly, the full set of states is typically not explicitly provided but instead can be derived
from the initial state by repeated application of the rules of the construction.

Using this definition, it is possible to represent the step-by-step behavior of an agent in a construction or gadget, as a sequence of states, each of the form $(q, \ell, s) \in (Q \times L \times S)$.
A particular state $(q, \ell, s)$ is considered \emph{observable} if $\ell$ is a location where the agent can enter and exit and is considered \emph{unobservable} otherwise.
For a gadget, the agent can enter and exit at all $L$, but a construction an agent can only enter and exit in $P \subset L$.

\begin{definition}[Trace]
A \emph{trace}, $T$, is a sequence of states, written explicitly as
\[(q_1, \ell_1, s_1) \rightarrow (q_2, \ell_2, s_2) \rightarrow \cdots \rightarrow (q_n, \ell_n, s_n)\]
with two conditions.
First, there is the initial condition $q_1 := q_{\text{start}}$.
Next, for all steps $i$, $((q_i, \ell_i, s_i) \rightarrow (q_{i+1}, \ell_{i+1}, s_{i+1}))$ must either be a transition in $\Delta$ or $q_i = q_{i+1}$ and both states are observable (emulating movement outside the construction/gadget).
\end{definition}
Note that the state $(q_{\text{start}}, \ell, s)$ is considered a valid trace and that traces can be infinite.
A trace of a gadget consists only of observable states, while a trace of a construction can consist of both.
To formally connect the two, we define a restricted view into a trace, called an observable transition sequence.

\begin{definition}[Observable transition sequence]
Given a trace 
\[T = (q_1, \ell_1, s_1) \rightarrow (q_2, \ell_2, s_2) \rightarrow \cdots \rightarrow (q_n, \ell_n, s_n)\]
its \emph{observable transition sequence} is a word over the alphabet $\Sigma = (L \times S) \to (L \times S)$ constructed by compressing unobservable transitions and only considering all observable states.
This is formalized in the following description which operates on sub-traces.

Let $i_1 < i_2 < \cdots < i_m$ be the ordered list of all indices such that $(q_{i_j}, \ell_{i_j}, s_{i_j})$ is observable.
We define the observable transition sequence as: $T^* = w_1 w_2 \cdots w_{m-1}$
where $w_j := ((\ell_{i_j}, s_{i_j}) \to (\ell_{i_{j+1}}, s_{i_{j+1}}))$
if the sub-trace
\[(q_{i_j}, \ell_{i_j}, s_{i_j}) \rightarrow (q_{i_j+1}, \ell_{i_j+1}, s_{i_j+1}) \rightarrow \cdots \rightarrow (q_{i_{j+1}}, \ell_{i_{j+1}}, s_{i_{j+1}})\]
satisfies the following two conditions:
\begin{enumerate}
  \item All intermediate states $(q_k, \ell_k, s_k)$ for $i_j < k < i_{j+1}$ are \emph{unobservable}.
  \item Every transition in the sub-trace is in $\Delta$, that is
  \[\forall k \in \{i_j, \ldots, i_{j+1} - 1\}, ((q_k, \ell_k, s_k) \to (q_{k+1}, \ell_{k+1}, s_{k+1})) \in \Delta.\]
\end{enumerate}
and $w_j$ is $\epsilon$, the empty character, otherwise.
\end{definition}
When defined this way, an observable transition sequence abstracts away the state of a construction or gadget.
Let $T_X$ denote the a trace of a construction or gadget $X$ and let $\mathcal{L}(X)$ to denote the language of observable transition sequences of $X$.
Given this setup, one can finally formally define correspondence between a construction and a gadget using the language of observational equivalence.

\begin{definition}[Observational equivalence]
\label{def:obseq}
Let $C = (Q_C, q^C_{\text{start}}, L_C, P_C, \Delta_C)$ be a construction and
$G = (Q_G, q^G_{\text{start}}, L_G, \Delta_G)$ be a gadget.
We say that the construction $C$ is \emph{observationally equivalent} to the gadget $G$, written $C \equiv_{\text{obs}} G$, if
\begin{enumerate}
    \item Their sets of observable locations are the same $P_C = L_G$ and
    \item The languages of their observable transition sequences are exactly identical $\mathcal{L}(C) = \mathcal{L}(G)$.
\end{enumerate}
\end{definition}
These conditions ensure that the construction respects the gadget's transition semantics and thus does not exhibit any observable behavior that is disallowed by the gadget.
They also ensure that the construction is expressive enough to realize every behavior that the gadget permits, at least at the observable level.
Together, these conditions guarantee that the construction and gadget are indistinguishable from the perspective of an external observer who can only see entry and exit at ports.
If these conditions hold, we say that the construction implements the gadget specification.

\begin{remark}
\label{rem:obseqweird}
As a technical point, note that this definition is strictly a formalization of single-agent, perfect information motion-planning problems.
A particular state in the gadget can represent a superposition of states in a corresponding construction,
and there can be states in a construction that do not map to states in an observationally equivalent gadget.
An example of these counterintuitive behaviors is shown in Figure~\ref{fig:odd}.

This behavior implicitly appears in prior works, emerges from the definition of observational equivalence, and can be explained in terms of agent strategy.
Note that understanding this nuance is not strictly required to understand Section~\ref{s:proof}.
\begin{figure}
  \centering
  \begin{minipage}[c]{0.19\textwidth}
    \centering
    \includesvg[width=\linewidth]{svgs/odd/odd1.svg}
    \caption*{(1)\vspace{1em}}
  \end{minipage}
  \begin{minipage}[c]{0.19\textwidth}
    \centering
    \includesvg[width=\linewidth]{svgs/odd/odd2.svg}
    \caption*{(2)\vspace{1em}}
  \end{minipage}
  \begin{minipage}[c]{0.19\textwidth}
    \centering
    \includesvg[width=\linewidth]{svgs/odd/odd3.svg}
    \caption*{(3)\vspace{1em}}
  \end{minipage}
  \begin{minipage}[c]{0.19\textwidth}
    \centering
    \includesvg[width=\linewidth]{svgs/odd/odd4.svg}
    \caption*{(4): start\vspace{1em}}
  \end{minipage}
  \begin{minipage}[c]{0.19\textwidth}
    \centering
    \includesvg[width=\linewidth]{svgs/odd/odd5.svg}
    \caption*{(5)\vspace{1em}}
  \end{minipage}
  \begin{minipage}[c]{0.47\textwidth}
      \begin{tikzpicture}[gadget]
        \node[state, initial]    (0) {(4)};
        \node[state]             (1) [right of=0, node distance=4cm] {(2)};
        \path (0) edge[loop above] node {$A \leftrightarrow C$} (0)
            edge[bend left, above] node {$\begin{aligned}B \to B\\B \to C\end{aligned}$} (1)
        (1) edge[bend left, below] node {$\begin{aligned}A \to A\\A \to C\end{aligned}$} (0)
            edge[loop above] node {$B \leftrightarrow C$} (1);
      \end{tikzpicture}
  \end{minipage}
  \begin{minipage}[c]{0.5\textwidth}
      \begin{tikzpicture}[gadget]
        \node[state]             (2) {};
        \node[state, initial]    (0) [above left of=2, node distance=4cm, yshift=-1cm] {(4)};
        \node[state]             (1) [above right of=2, node distance=4cm, yshift=-1cm] {(2)};
        \path (0) edge[loop above] node {$A \leftrightarrow C$} (0)
            edge[bend left] node {$B \to C$} (1)
            edge[bend right, below left] node {$B \to B$} (2)
        (1) edge[above] node {$A \to C$} (0)
            edge[loop above] node {$B \leftrightarrow C$} (1)
            edge[bend left, below right] node {$A \to A$} (2)
        (2) edge[above right] node {$A \leftrightarrow C$} (0)
            edge[above left] node {$B \leftrightarrow C$} (1);
      \end{tikzpicture}
  \end{minipage}
  \caption{Observable states of a Push-1 construction (top), two gadgets (left) and (right) with trivial transitions omitted.
  Note that the construction and gadgets all satisfy observational equivalence; however,
  in both cases, construction states $1$, $3$, and $5$ are not reflected in any gadget states.
  In the second gadget, there is an unlabeled state corresponding to the possibility that the construction is either in state $(2)$ or in state $(4)$.
  When the agent is in this state and walks to or from $C$ and another port,
  the transition collapses the superposition of these states into either $(2)$ or $(4)$.}
  \label{fig:odd}
\end{figure}
 
To explain succinctly, suppose the agent enters a construction in state $q$ at port $\ell$ and has the option to exit at port $\ell'$ and change the state to either $q_a$ or $q_b$.
In the construction, the agent must explicitly make this choice; however, in a gadget, the agent can defer or even ignore the choice.

If the observable transitions from these two states are distinct, that is there exists a $(q_a, \ell, s) \rightarrow (q', \ell', s')$ without a corresponding $(q_b, \ell, s) \rightarrow (q', \ell', s')$ and vice versa, then the agent can enter a gadget state which corresponds to the construction being in an ambiguous state (either $q_a$ or $q_b$) until it takes one of these transitions.
This explains why a particular state in the gadget can represent a superposition of states in a corresponding construction.

If for every observable transition $(q_a, \ell, s) \rightarrow (q', \ell', s')$,
there exists a an observable transition $(q_b, \ell, s) \rightarrow (q', \ell', s')$, then the gadget never needs to contain the state $q_a$ since it is weakly dominated by $q_b$.
This explains why there can be states in a construction that do not map to states in an observationally equivalent gadget.
\end{remark}

\subsection{Gadget synthesis}

Given a construction $C$ and a gadget $G$,
suppose a program could construct non-deterministic finite automata (NFA) $\mathcal{A}_C$ and $\mathcal{A}_G$ which recognize $\mathcal{L}(C)$ and $\mathcal{L}(G)$ respectively.
In that case, the same program could also verify that these automata recognize the same languages via standard techniques (i.e. by checking that the symmetric difference recognizes an empty language),
and thus it could prove the observational equivalence of $C$ and $G$.

Additionally, suppose that a program was only provided $C$.
If the process for constructing an automaton from a gadget was bidirectional, meaning that it was possible to translate from gadget to automata and from automata to gadget, then
the following 3-step procedure would suffice for synthesizing a minimal (deterministic) $G$ from $C$ which is correct \emph{by construction}.
\begin{enumerate}
    \item Given $C$, generate a non-deterministic finite automaton $\mathcal{A}_C$ that recognizes $\mathcal{L}(C)$.
    \item Determinize and minimize $\mathcal{A}_C$ as $\mathcal{A}_G$.
    \item Convert the automaton $\mathcal{A}_G$ into a gadget $G$.
\end{enumerate}
Note that determinizing, minimizing, and verifying the equivalence of automata are all standard procedures~\cite[Chapter~1]{sipser13intro}. Here we provide procedures for steps 1 and 3.

\paragraph{From construction to automaton}
To generate an NFA $\mathcal{A}_C$ that recognizes $\mathcal{L}(C)$, one can perform a search over the space of observable transitions.
The key idea is that states of $\mathcal{A}_C$ (with the exception of a single trap state) will correspond to observable states of the construction $C$.
A transition between $q$ and $q'$ in $\mathcal{A}_C$ will exist iff for $\ell, \ell' \in P_C$, there is a sequence of transitions going from $(q, \ell, s)$ to $(q', \ell', s')$ in which no intermediate state is observable.
The algorithm, formally provided below, uses a search on observable states.

\begin{enumerate}
    \item Initialize a queue and a seen set with all initial observable configurations of $C$, i.e., all states of the form $(q_{\text{start}}, \ell, s)$ where $\ell \in P_C$.
    \item Initialize the NFA $\mathcal{A}_C = (Q_{\mathcal{A}}, \Sigma, \delta_{\mathcal{A}}, q_0, F)$ where:\begin{itemize}
        \item The alphabet is $\Sigma = (P_C \times S) \to (P_C \times S)$.
        \item The set of states $Q_{\mathcal{A}}$ begins as the pair $\{q_{\text{start}}, q_{\text{halt}}\}$.
        \item The transition function $Q_{\mathcal{A}} \times \Sigma \to Q_{\mathcal{A}}$, by default, maps all state-symbol pairs to $q_{\text{halt}}$.
        \item the start state $q_0$ is simply $q_{\text{start}}$.
        \item The set of accepting states $F$ begins as the singleton set $\{q_{\text{start}}\}$.
    \end{itemize}
    \item Until the queue is empty\begin{itemize}
        \item Pop the next observable configuration $(q, \ell, s)$ from the queue:
        \item Perform a search (breadth-first or depth-first) through transitions in $\Delta_C$, starting at $(q, \ell, s)$ to identify all $(q', \ell', s')$ where $\ell' \in P_C$ reachable only by a sequence of unobservable states.
        \item For all $(q', \ell', s')$, if $q'$ is not in $Q_{\mathcal{A}}$, add it and mark it as an accepting state.
        \item For all $(q', \ell', s')$, add $(q \times ((\ell, s) \to (\ell', s'))) \to q'$ to $\delta_{\mathcal{A}}$.
        \item For all $(q', \ell', s')$ where $\ell' \in P_C$ not in seen, add them to the queue.
    \end{itemize}
\end{enumerate}
By definition, the resulting automaton recognizes $\mathcal{L}(C)$ as every valid observable transition corresponds 1-to-1 with a transition between states in $Q_{\mathcal{A}} \setminus \{q_{\text{halt}}\}$.

Now this resulting automaton can be determinized and minimized without changing the language it recognizes.
Once the automaton has finished processing, it needs to be turned back into a gadget.

\paragraph{From automaton to gadget}

Converting from automaton to gadget is relatively straightforward.
Given $\mathcal{A}_G = (Q_{\mathcal{A}}, \Sigma, \delta_{\mathcal{A}}, q_0, F)$.
\begin{itemize}
    \item Let $Q_G = Q_{\mathcal{A}} \setminus \{q_{\text{halt}}\}$ be the set of gadget states.
    \item Let $q^G_{\text{start}} = q_0$.
    \item Let $L_G = P_C$ where $P_C$ is contained in $\Sigma$.
    \item Let $\Delta_G$ contain $((q, \ell, s) \to (q', \ell', s'))$ iff $q \times ((\ell, s), (\ell', s')) \to q' \in \delta_{\mathcal{A}}$ (and $q$ and $q'$ are not $q_{\text{halt}}$).
\end{itemize}
This process produces a gadget whose observable behavior matches that of the automaton exactly, and hence (by design) also matches the observable behavior of the original construction $C$.

This synthesis guarantees that $G$ is the most succinct deterministic abstraction of $C$'s observable behavior, making it useful both for correctness proofs and for designing higher-level gadgets from lower-level implementations.

Checkability and post-selection can then be enforced in the ways described in prior works~\cite{ani22pushing}.

\subsection{Systems of gadgets are gadgets}

In the prior exposition, there was a clear distinction between constructions, which potentially contain unobservable internal behavior, and gadgets, which are defined to exclusively contain observable behavior.
However, there was no discussion of the behavior or verification of systems of gadgets (Definition~\ref{def:sys}).
This is because a system of gadgets, with some auxiliary information, is a construction.

Let $H$ be the gadgets in the system and $E$ be the set of edges connecting locations in the system.
The set of states $Q$ is the cartesian product of the states of all gadgets, $Q := \prod_{G_i \in H} Q_{G_i}$.
The initial state $q_{\text{start}}$ is the cartesian product of initial states, $q_{\text{start}} := \prod_{G_i \in H} q^{G_i}_{\text{start}}$.
The set of locations $L$ is the union of locations in all gadgets, $L := \bigcup_{G_i \in H} L_{G_i}$.
The set of ports $P$ is specified with auxiliary information as an arbitrary subset of $L$.
The set of transitions $\Delta$ includes two types of transitions.
\begin{enumerate}
    \item \textbf{Transitions within gadgets.}  
    For any gadget $G_i \in H$, if $((q_i, \ell, s), (q_i', \ell', s')) \in \Delta_{G_i}$, then
    \[(((q_1, \dots, q_i, \ldots, q_n), \ell, s) \rightarrow
    ((q_1, \dots, q_i', \ldots, q_n), \ell', s')) \in \Delta.\]
    That is, the agent moves within a single gadget according to that gadget's semantics, only updating that gadget’s internal state and the agent's location and state.
    \item \textbf{Traversals between gadgets.}  
    For any edge $(\ell, \ell') \in E$,
    \[(((q_1, \ldots, q_n), \ell, s) \rightarrow
    ((q_1, \ldots, q_n), \ell', s)) \in \Delta.\]
    This models movement of the agent between gadgets, without changing the state of any gadget.
\end{enumerate}
In this light, the techniques for synthesizing gadgets from constructions, immediately apply to synthesizing more complex gadgets from simpler systems of gadgets.
It is similarly possible to intermingle constructions and gadgets in a system, and then to apply the techniques above for synthesis without first individually converting each construction into a gadget.

\subsection{Implementation}
We implemented the generic gadget synthesis algorithm capable of working with arbitrary user-specified constructions
and encoded the ruleset of Push-1 with careful attention to handle passing blocks between gadgets.
This system, which we call \sysname, and the Push-1 rules were implemented in 1,380 lines of Rust with multi-threading.

 \section{Push-1 is PSPACE-complete}
\label{s:proof}

We build a variant of the checkable self-closing door design from \textit{Pushing Blocks via Checkable Gadgets}~\cite{ani22pushing} using two diodes, one 1-toggle, a precursor, and various other smaller gadgets to ensure the check-path.
The remainder of this section walks through the gadgets in order of complexity, presenting the constructions, the resulting synthesized gadgets produced by \sysname, and any notable features.
Some of these constructions gadgets are adapted from designs in \textit{Pushing Blocks via Checkable Gadgets}.
We present all gadgets, new and old, which contribute to the final self-closing door, distinguishing between those that are novel and those for which only the formal verification is novel.

Ports are denoted by capital letters for reference.
For ports that are configured to only allow the agent to enter if it is not pushing a block and/or do not allow the agent to exit with a block, we omit the agent's state for readability. Additionally, we omit trivial transitions in which the agent enters and exits the same port without changing the state of the gadget.

For clarity in visualizing the mechanics of the constructions, sections of the grid with permanently immobile blocks are denoted differently from other, movable blocks.
Recall that these immobile blocks are immobile purely by their positioning, not because they differ
from the other blocks in any way.
In all cases in this paper, the blocks will be part of $2 \times 2$ squares; however,
there are other layouts that result in permanently immobile blocks.

\begin{figure}
  \centering
  \begin{minipage}[c]{0.99\textwidth}
    \centering
    \includesvg[width=\linewidth]{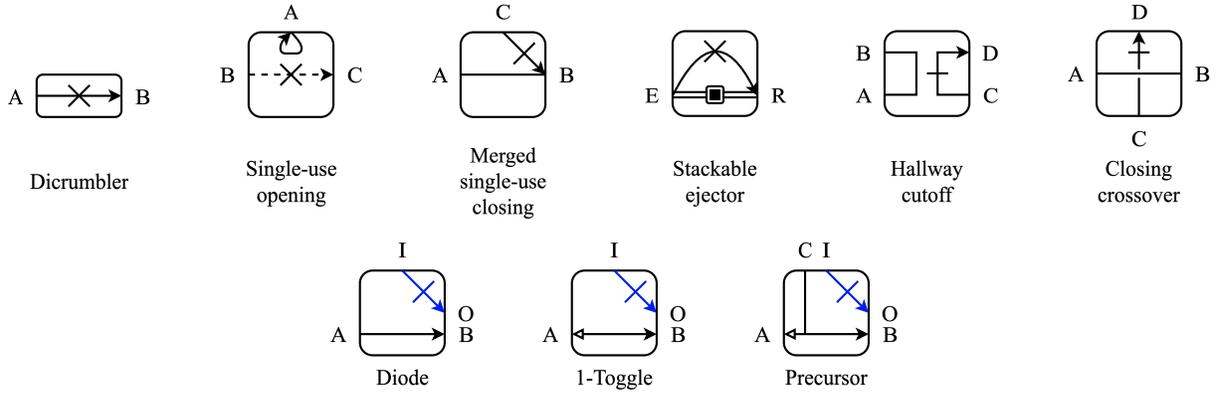}
  \end{minipage}
  \caption{Table of symbols for sub-gadgets contained in the proof.}
  \label{fig:guide}
\end{figure}
 
Figure~\ref{fig:guide} contains a table of gadgets and their symbols when used in systems of gadgets.
Some of these symbols are standard, some are introduced by us in the style of standard symbols.
These are used throughout the proof of hardness of Push-1 for indicating direction, orientation, and behavior at a glance within systems of gadgets.

\paragraph{Dicrumbler}

\begin{figure}
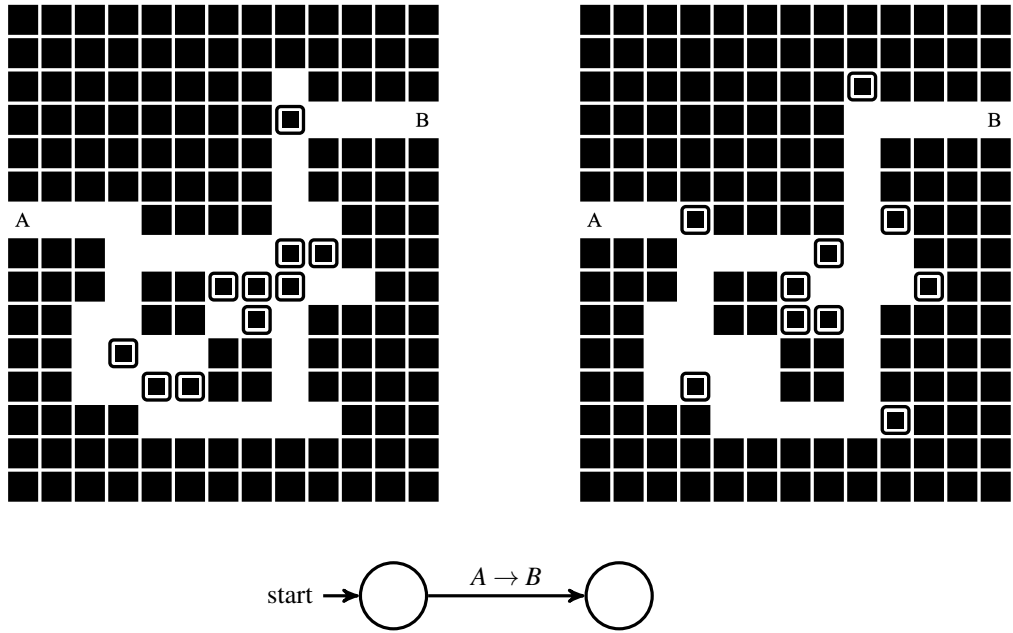

  \centering
  \begin{minipage}[c]{0.35\textwidth}
    \centering
    \includesvg[width=\linewidth]{svgs/sd}
  \end{minipage}
  \hspace{0.1\textwidth}
  \begin{minipage}[c]{0.35\textwidth}
    \centering
    \includesvg[width=\linewidth]{svgs/sd-done}
  \end{minipage}\\
  \vspace{2em}
  \hspace{-4em}\begin{minipage}[c]{0.4\textwidth}
    \centering
    \begin{tikzpicture}[gadget]
      \node[state, initial]    (0)                    {};
      \node[state]             (1) [right of=0]       {};
      \path (0) edge  node {$A \to B$} (1);
    \end{tikzpicture}
  \end{minipage}
  \caption{Dicrumbler construction (left), construction after transition from $A \to B$ (right), and gadget (bottom).
  Observe that in the second construction state, the path between $A$ and $B$ is blocked permanently by an immobile block near $A$.
  There are two blocks which can still be pushed by the agent after entering $B$; however, neither of these blocks can exit the construction.
  The blocks which were movable in the first construction are highlighted in second construction, even if most of them are now permanently immobile.}
  \label{fig:sd}
\end{figure}

A dicrumbler, depicted in Figure~\ref{fig:sd}, is a gadget which only allows the agent to enter $A$ and then exit $B$ exactly once.
An agent cannot enter $B$ and afterwards it is fully closed to non-trivial transitions.
The design below does not allow for the agent to bring blocks into the gadget.
This design is simpler and more compact than prior work, and it demonstrates the conversion from Push-1 construction to gadget.

Additionally, there are several moves the agent can make, after entering $A$
which fully block the path to $B$.
However, there is no transition $A \to A$ in the gadget to a locked state of this form.
This highlights an important subtlety that emerges from Definition~\ref{def:obseq} and was highlighted in Remark~\ref{rem:obseqweird},
which was implicitly used but not explicitly described in prior works.

\pagebreak

\paragraph{Single-use opening}

\begin{figure}
  \centering
  \begin{minipage}[c]{0.7\textwidth}
    \centering
    \includesvg[width=\linewidth]{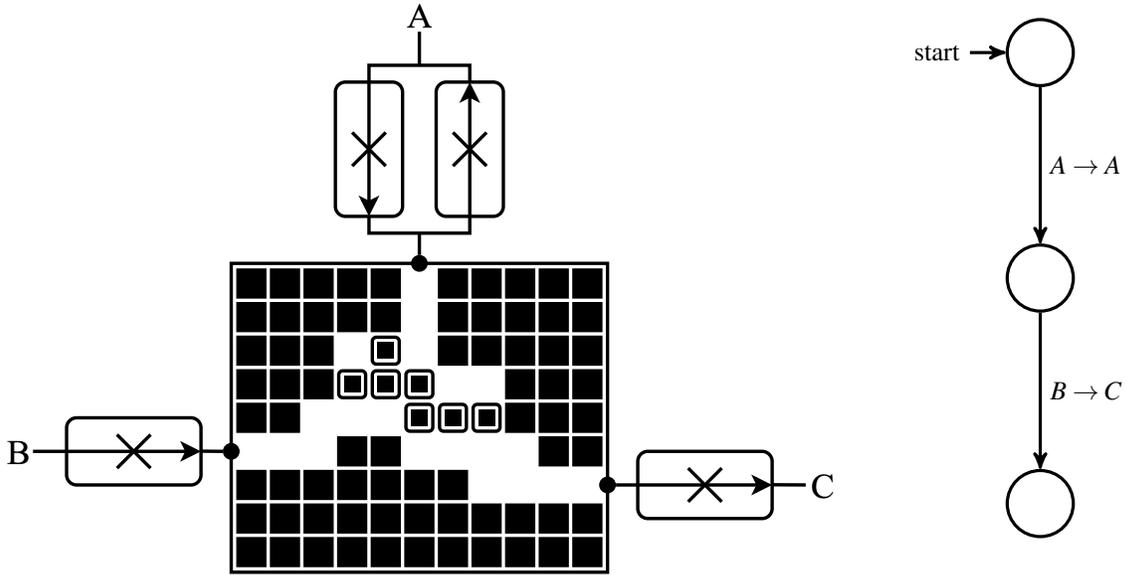}
  \end{minipage}
  \begin{minipage}[c]{0.25\textwidth}
    \centering
    \begin{tikzpicture}[gadget]
      \node[state, initial]    (0)                    {};
      \node[state]             (1) [below of=0]       {};
      \node[state]             (2) [below of=1]       {};
      \path (0) edge  node {$A \to A$} (1);
      \path (1) edge  node {$B \to C$} (2);
    \end{tikzpicture}
  \end{minipage}
  \caption{Single-use opening construction (left) and gadget (right).}
  \label{fig:so}
\end{figure}

The single-use opening, depicted in Figure~\ref{fig:so}, is a hallway from $B$ to $C$, which is closed at first, then opened by entering and exiting at $A$, and then closed permanently after it is traversed once.
Note that the construction here is nearly identical to prior work.
Observe that the construction is depicted as combining a raw Push{-}1 construction with 4 dicrumblers.
It is, of course, possible to diagram every block of the construction; however, the presentation as a system of gadgets and constructions allows for modular reasoning.

\paragraph{Merged single-use closing}

\begin{figure}
  \centering
  \begin{minipage}[c]{0.5\textwidth}
    \centering
    \includesvg[width=\linewidth]{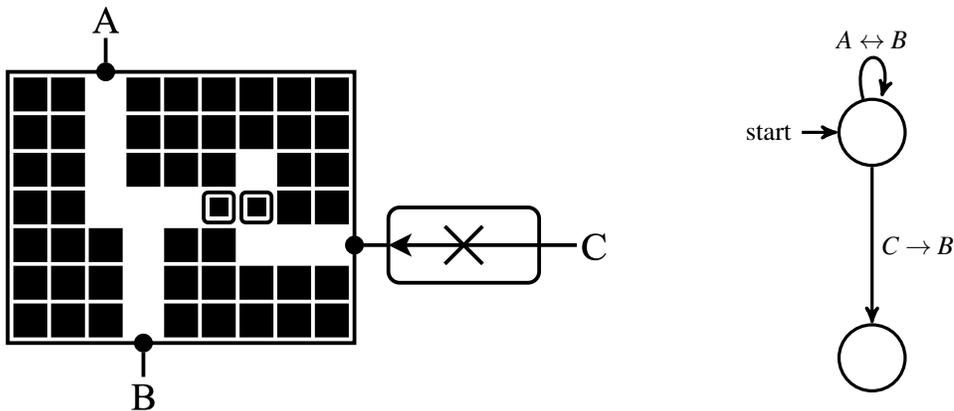}
  \end{minipage}
  \begin{minipage}[c]{0.35\textwidth}
    \centering
    \begin{tikzpicture}[gadget]
      \node[state, initial]    (0)                    {};
      \node[state]             (1) [below of=0]       {};
      \path (0) edge[loop above]  node {$A \leftrightarrow B$} (0);
      \path (0) edge  node {$C \to B$} (1);
    \end{tikzpicture}
  \end{minipage}
  \caption{Merged single-use closing construction (left) and gadget (right).}
  \label{fig:msc}
\end{figure}

\textit{Pushing Blocks via Checkable Gadgets}'s~\cite{ani22pushing} design for the 
merged single-use closing (Figure~\ref{fig:msc})
is reproduced here without fixed blocks and formally verified.

This gadget allows an agent to travel between $A$ and $B$ until a transition of the form $C \to B$, at which point no more non-trivial transitions are permitted.

\pagebreak

\paragraph{Stackable ejector and $k$-block ejector}

The stackable ejector allows the agent to travel from $A \to B$ once before forcing the agent to eject a block out $A$ (or $B$) to open the passageway between $A$ and $B$ for future travel.
By combining several of these in series with a single-use opening at the end,
one gets a $k$-block ejector,
a gadget which requires an agent to push $k$ blocks out of it to traverse from $A$ to $B$.
These are the first gadgets in this list to involve the agent's state, thus motivating Definition~\ref{def:gad2}.

These constructions specifically rely on the fact that the agent can only push one block at a time.
When traveling from $A \to B$, this forces the agent to take the dicrumbler.
When attempting to travel from $B \to A$, prior to ejection, this prevents the agent from bringing in a block.

\begin{figure}
  \centering
  \begin{minipage}[c]{0.65\textwidth}
    \centering
    \includesvg[width=\linewidth]{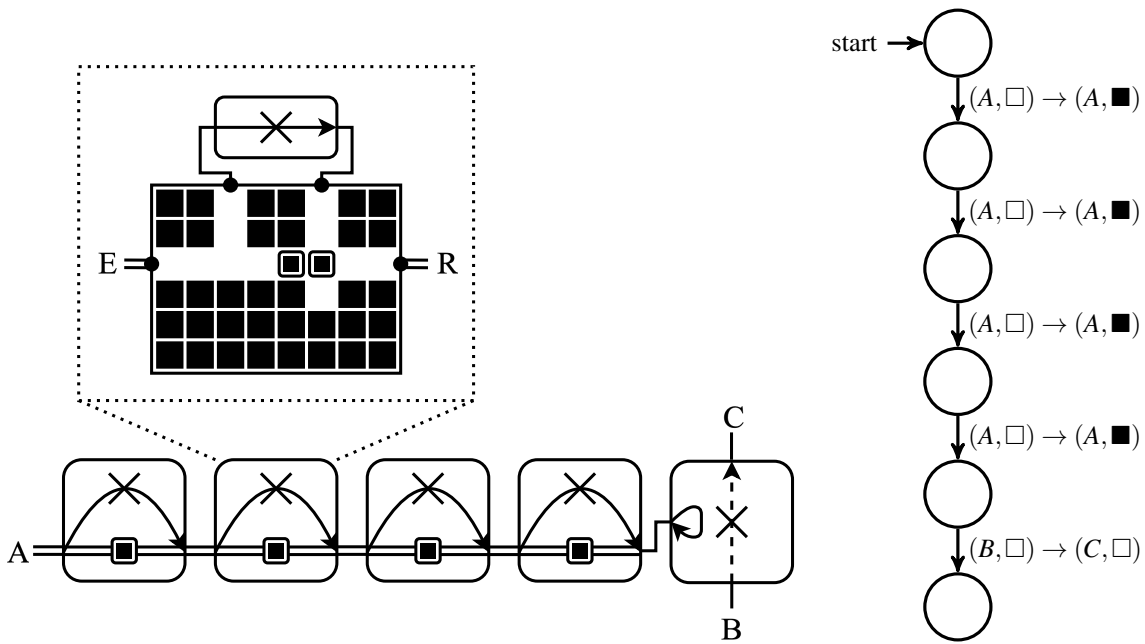}
  \end{minipage}
  \begin{minipage}[c]{0.3\textwidth}
    \centering
    \begin{tikzpicture}[gadget, node distance=1.5cm]
      \node[state, initial]    (0)                    {};
      \node[state]             (1) [below of=0]       {};
      \node[state]             (2) [below of=1]       {};
      \node[state]             (3) [below of=2]       {};
      \node[state]             (4) [below of=3]       {};
      \node[state]             (5) [below of=4]       {};
      \path (0) edge  node {$(A, \square) \to (A, \blacksquare)$} (1);
      \path (1) edge  node {$(A, \square) \to (A, \blacksquare)$} (2);
      \path (2) edge  node {$(A, \square) \to (A, \blacksquare)$} (3);
      \path (3) edge  node {$(A, \square) \to (A, \blacksquare)$} (4);
      \path (4) edge  node {$(B, \square) \to (C, \square)$} (5);
    \end{tikzpicture}
  \end{minipage}
  \caption{4-block ejector construction with highlighted stackable ejector (left) and simplified gadget (right).
  This gadget omits transitions where a block is pushed into the ejector and omits all subsequent states for clarity.
  In a transition, $\square$ indicates that the agent is stepping and $\blacksquare$ indicates that it is pushing a block.
  Thus $(A, \square) \to (A, \blacksquare)$ should be read as, ``the agent enters location $A$ without a block and subsequently pushes a block out of $A$.''
  In the stackable ejector construction, the agent enters $E$, and needs to collapse the dicrumbler and push one block down, to go to $R$.
  To return to $E$ from $R$, the agent is required to eject the block out $E$,
  after which the hallway is open.
  The 4-block ejector forces the agent to repeat this 4 times, ejecting blocks through prior stackable ejectors.}
  \label{fig:e4}
\end{figure}

The $k$-block ejector gracefully handles blocks being pushed into it, preventing an agent from storing ejected blocks back in the body of the construction.
We have opted to include the construction for the stackable ejector and the full construction and partial (simplified) gadget representations
for a $4$-block ejector in Figure~\ref{fig:e4}.
Gadgets of this form contain a larger number of states,
most of which handle blocks being brought back into ejectors.
Most of these states will never be encountered when these gadgets are used as part of larger constructions.

These ejectors are central to the design of the most complex gadgets as they
force more complex block manipulation by agents.

\paragraph{Hallway cutoff and closing crossover}

\begin{figure}
  \centering
  \begin{minipage}[c]{0.5\textwidth}
    \centering
    \includesvg[width=\linewidth]{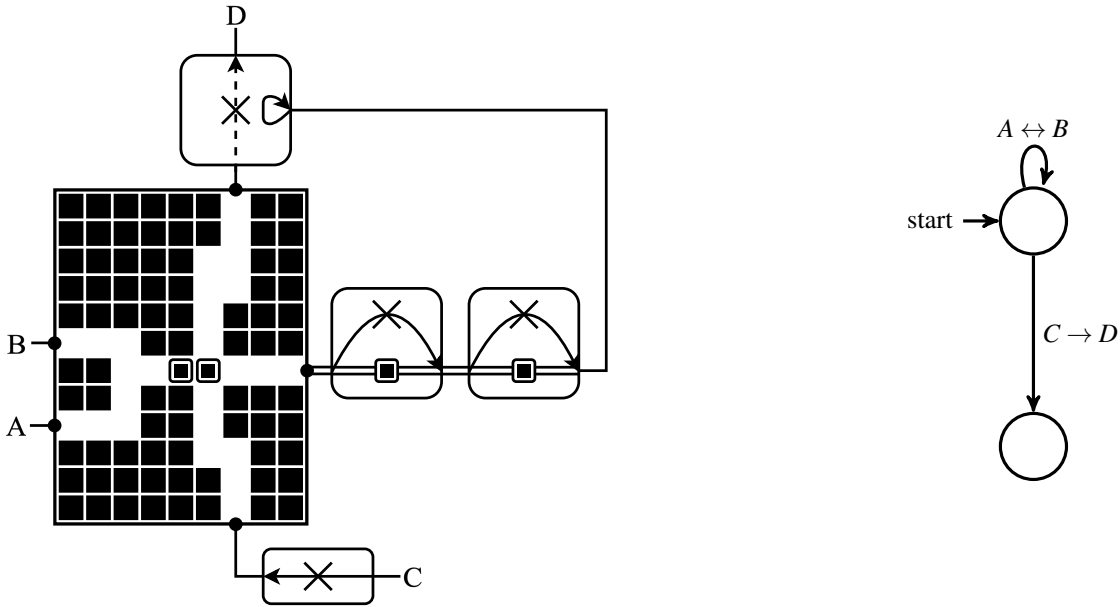}
  \end{minipage}
  \hfill
  \begin{minipage}[c]{0.35\textwidth}
    \centering
    \begin{tikzpicture}[gadget]
      \node[state, initial]    (0)                    {};
      \node[state]             (1) [below of=0]       {};
      \path (0) edge[loop above]  node {$A \leftrightarrow B$} (0);
      \path (0) edge  node {$C \to D$} (1);
    \end{tikzpicture}
  \end{minipage}
  \caption{Hallway cutoff construction (left) and gadget description (right).}
  \label{fig:hc}
\end{figure}

\begin{figure}
  \centering
  \begin{minipage}[c]{0.6\textwidth}
    \centering
    \includesvg[width=\linewidth]{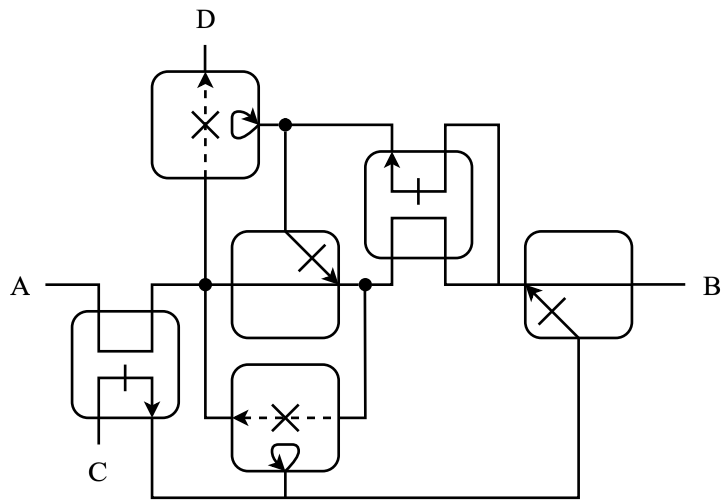}
  \end{minipage}
  \caption{Closing crossover construction as a system of gadgets. Its gadget description is identical to that of Figure~\ref{fig:hc}.}
  \label{fig:cx}
\end{figure}

The hallway cutoff and closing crossover, depicted in Figures~\ref{fig:hc}~and~\ref{fig:cx} respectively, allow for the agent to freely transition between $A$ and $B$ until a transition of the form $C \to D$ occurs, at which point no other non-trivial transitions are permitted.
They are distinguished topologically.
The former has the $A \to B$ and $C \to D$ hallways parallel to each other in the grid,
while the latter has the two hallways perpendicular, crossing over each other.
They are both useful to permanently prevent access to areas of a gadget and to ensure the resulting self-closing door is planar.

The hallway cutoff explicitly uses two stackable ejectors in series.
The agent is forced to make room for these ejected blocks, and in the process, is forced to close
off the path between $A$ and $B$.
This provides an example of how ejectors can force non-local behavior, specifically forcing a block not contained in the ejector to be moved into a specific location.
This can be framed as the ejector checking that some other block is in a correct location,
and provides a preview of the techniques used for the subsequent complex gadgets.

The closing crossover takes inspiration from prior weak-closing crossover designs.
It is possible to add two hallway cutoffs to a prior weak-closing crossover design~\cite{ani22pushing}, turning it into a closing crossover; however, we demonstrate that the whole design can be made conceptually simpler.
In prior works, the weak closing crossover was called \emph{leaky}, as it allowed for the agent to transition from $C \to B$ (in addition to $C \to D$) when entering the crossover path.
This was because the design could not seal off access to both $A$ and $B$ at the same time.
In contrast, the hallway cutoff allows for constructions that seal an arbitrary number of locations.

\pagebreak

\paragraph{Diode}

The diode is a gadget that allows for any number of transitions of the form $A \to B$ but strictly
disallows those of the form $B \to A$.
This diode is the first post-selected (checkable) gadget and implicitly uses ejections as part of the checking path.
Prior work~\cite{ani22pushing} was prevented from constructing this gadget in large part due to the need for their design to include $1 \times 1$ fixed-blocks.
We believe that the key to overcoming the need for fixed-blocks is block ejections.

\begin{figure}
  \centering
  \begin{minipage}[c]{0.3\textwidth}
    \centering
    \includesvg[width=\linewidth]{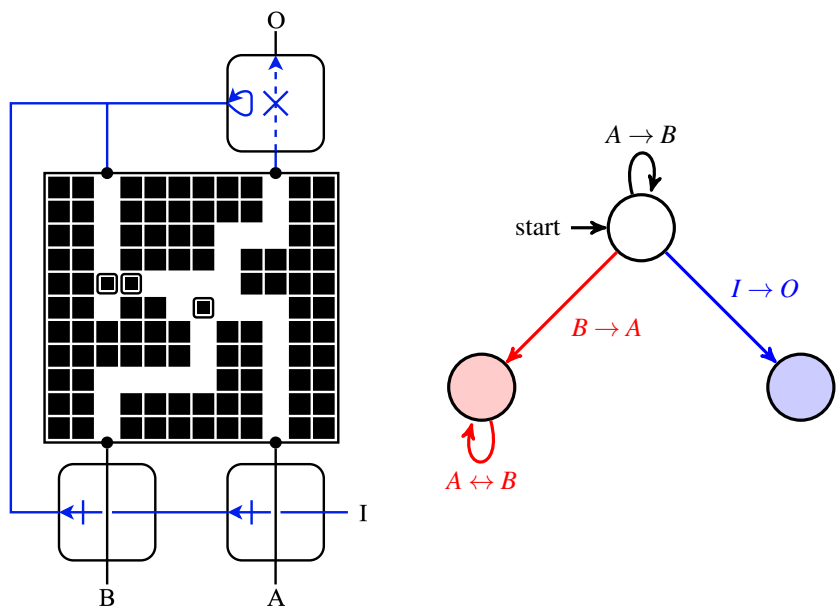}
  \end{minipage}
  \begin{minipage}[c]{0.4\textwidth}
    \centering
    \begin{tikzpicture}[gadget]
      \node[state, initial]    (0)                    {};
      \node[state, fill=red!20]             (1) [below left of=0]       {};
      \node[state, fill=blue!20]             (2) [below right of=0]       {};
      \path (0) edge[loop above]  node {$A \to B$} (0);
      \path (0) edge[draw=red, text=red]  node {$B \to A$} (1);
      \path (1) edge[loop below, draw=red, text=red]  node {$A \leftrightarrow B$} (1);
      \path (0) edge[draw=blue, text=blue]  node {$I \to O$} (2);
    \end{tikzpicture}
  \end{minipage}
  \caption{Checkable diode construction (left) and gadget prior to post-selection (right).
  The check transition $I \to O$ is denoted in blue, after which no other non-trivial traversals are allowed.
  In normal operation, the agent can travel from $A \to B$ repeatedly.
  While in the initial state, it is possible for the agent to travel from $B \to A$;
  however, this permanently prevents the checking traversal from succeeding.
  This breaking transition is denoted in red and removed in the process of post-selection.}
  \label{fig:diode}
\end{figure}

The diode is depicted in Figure~\ref{fig:diode}.
The main room of this diode is a $2 \times 3$ rectangle with a block on the lower row.
The normal behavior of this gadget is for the agent to enter $A$, loop around the block (potentially pushing it into the starting position again) and push it one tile to the right before continuing out $B$.

This diode is broken if this block is instead pushed into the left corner of the room or anywhere on the top row, as this allows for the agent to travel from $B$ to $A$.
Here, a single block is ejected into the top row on the check-path, effectively detecting and preventing both types of broken behavior. Additionally, as this will recur in subsequent gadgets, once the check-path is entered, it cannot be exited, and access in and out of non-checking path ports is fully cut using hallway cutoffs.

\paragraph{1-Toggle and precursor}

\begin{figure}
  \centering
  \begin{minipage}[c]{0.7\textwidth}
    \centering
    \includesvg[width=\linewidth]{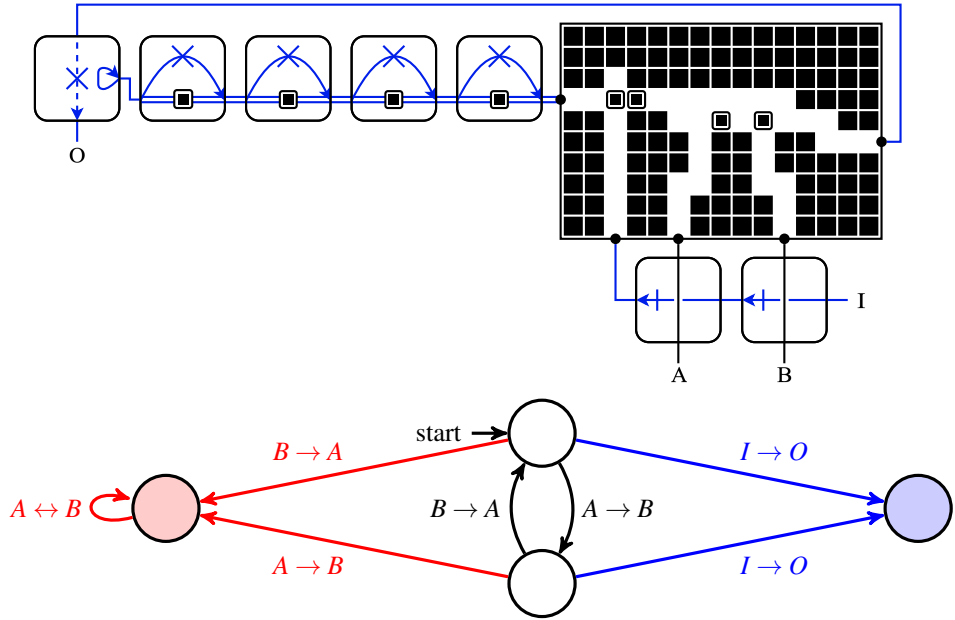}
  \end{minipage}
  \begin{minipage}[c]{0.75\textwidth}
    \centering
    \begin{tikzpicture}[gadget]
      \node[state, initial] (0) at (0,0) {};
      \node[state] (1) at (0,-2) {};
      \node[state, fill=red!20] (2) at (-5,-1) {};
      \node[state, fill=blue!20] (3) at (5,-1) {};
      \path (0) edge[bend left]  node {$A \to B$} (1);
      \path (0) edge[draw=red, text=red, above left]  node {$B \to A$} (2);
      \path (0) edge[draw=blue, text=blue]  node {$I \to O$} (3);
      \path (1) edge[bend left]  node {$B \to A$} (0);
      \path (1) edge[draw=red, text=red]  node {$A \to B$} (2);
      \path (1) edge[draw=blue, text=blue, below right]  node {$I \to O$} (3);
      \path (2) edge[loop left, draw=red, text=red]  node {$A \leftrightarrow B$} (2);
    \end{tikzpicture}
  \end{minipage}
  \caption{Checkable 1-toggle construction (above) and gadget prior to post-selection (below).
  The check transition $I \to O$ is denoted in blue, after which no other non-trivial traversals are allowed.
  In normal operation, the agent can travel from $A \to B$ and then from $B \to A$ as many times as required. If the agent tries to repeat the same transition two, for example, $A \to B$ and $A \to B$,
  it is possible, but it permanently prevents the check from succeeding.
  In normal operation, the agent enters from the bottom, shifts both blocks in the main room left or right by a single square and exits in the opposite direction at the bottom.}
  \label{fig:toggle}
\end{figure}
 \begin{figure}
  \centering
  \begin{minipage}[c]{0.7\textwidth}
    \centering
    \includesvg[width=\linewidth]{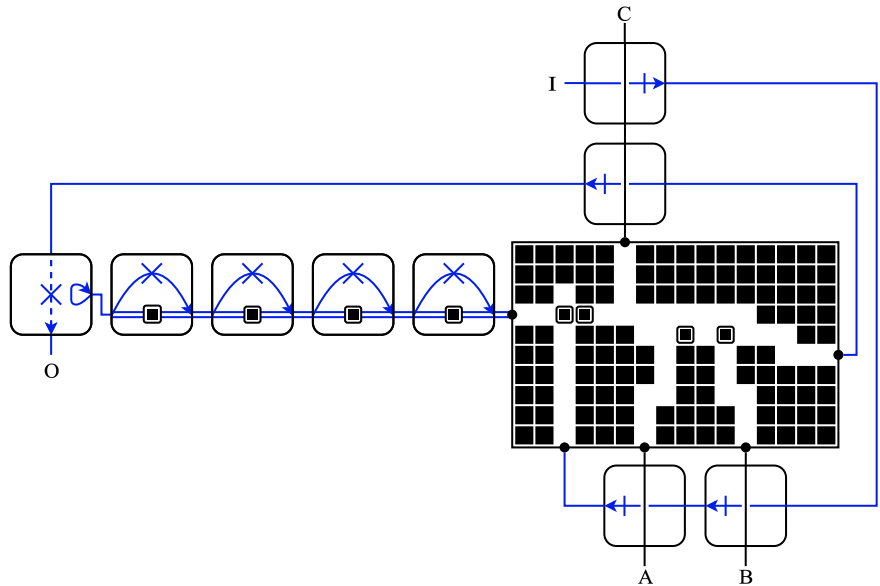}
  \end{minipage}
  \caption{Checkable precursor construction.
  This device acts similarly to the 1-toggle, with the addition of an entrance on the top, $C$, which allows for changing the state of the toggle.
  After entering $C$, the agent can exit $A$, $B$, or $C$, and from any entrance, the agent can exit $C$.
  The gadget is omitted, since it is a slight change to the one presented in Figure~\ref{fig:toggle}.
  The check transition is shown in blue.
  Note that after entering $C$, the agent can potentially exit $A$ or $B$, and from any entrance, the agent can exit $C$.
  When part of the checkable self-closing door, these transitions are restricted by additional post-selection.}
  \label{fig:pr}
\end{figure}
 
The 1-toggle is a gadget with parity.
In one state, the agent can travel $A \to B$.
In the other, the agent can travel $B \to A$.
Once the agent takes a transition of either type, the state changes to the opposite one.
The precursor behaves similarly; however, it has a hatch, $C$, that can be entered to toggle the state.
Note that both of these gadgets can be instantiated with opposite parity if required.

These constructions are virtually identical, but we present both for completeness in Figures~\ref{fig:toggle}~and~\ref{fig:pr}.
The main room of these constructions is a $2 \times 6$ rectangle that contains two blocks in the lower row.
For the correct functioning of these constructions, the blocks cannot be stored in the corners, pushed together in the middle, or moved onto the top row, as any one of these moves breaks the design.
The check-path consists of ejecting $5$ blocks into this main room in the top row and forces traveling from one end of the room to the other after it is nearly filled with blocks.
If the agent attempts any forbidden moves with the blocks in the gadget, the check-path cannot be completed.

The constraints and logic of this design were particularly complex.
After several broken attempts, this motivated the implementation of \sysname to discover flaws in earlier designs and to formally verify the final result.

\paragraph{Self-closing door}

Recall that the self-closing door is a gadget with an entrance, an exit, and a keyhole.
When the door is open, the agent can go from entrance to exit, closing the door in the process.
When the door is closed, the agent can reopen it by entering and exiting the keyhole.

\begin{figure}
  \centering
  \begin{minipage}[c]{0.55\textwidth}
    \centering
    \includesvg[width=\linewidth]{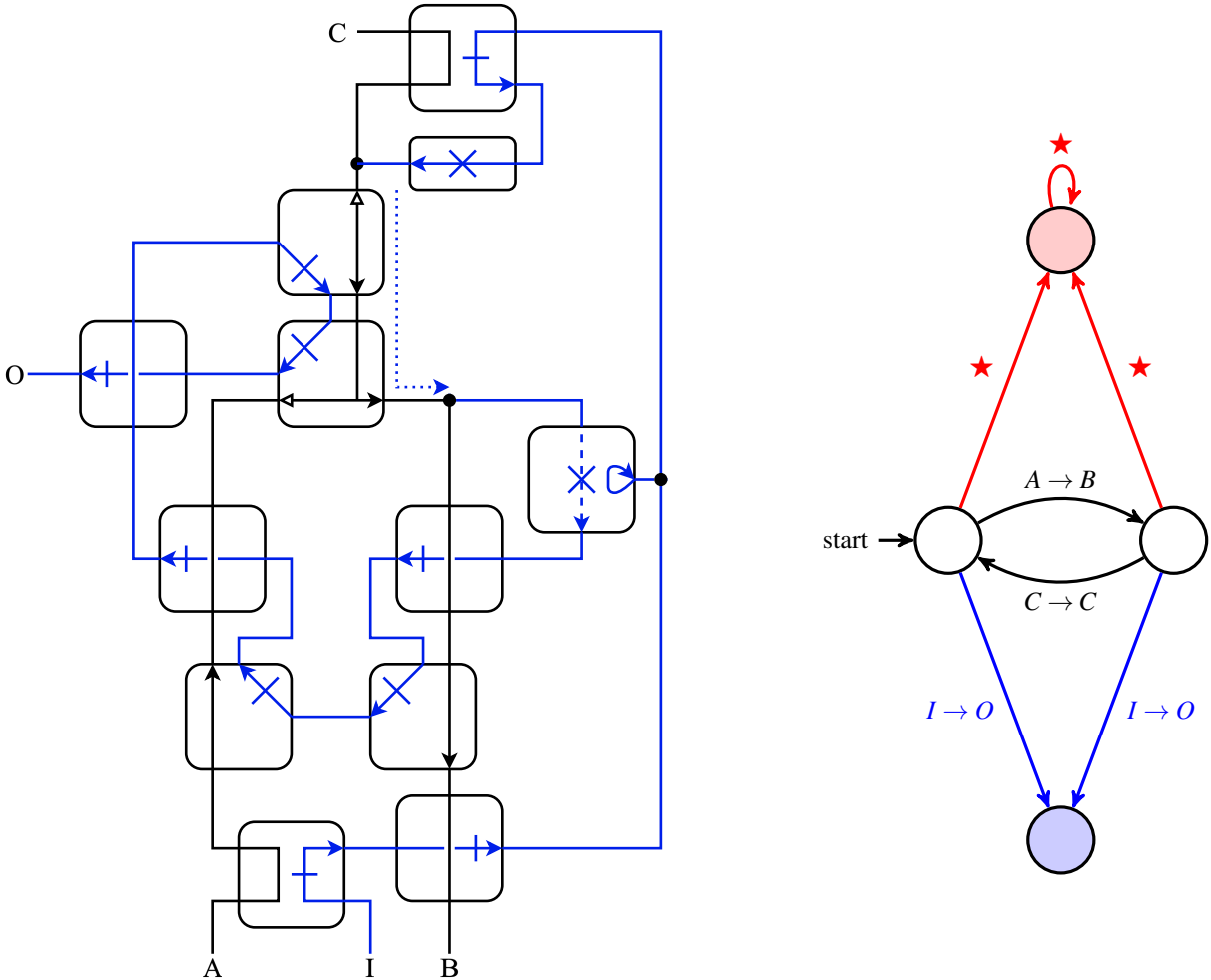}
  \end{minipage}
  \hfill
  \begin{minipage}[c]{0.35\textwidth}
    \centering
    \begin{tikzpicture}[gadget]
      \node[state, initial] (0) at (0,0) {};
      \node[state] (1) at (3,0) {};
      \node[state, fill=red!20] (2) at (1.5,4) {};
      \node[state, fill=blue!20] (3) at (1.5,-4) {};
      \path (0) edge[bend left]  node {$A \to B$} (1);
      \path (1) edge[bend left]  node {$C \to C$} (0);
      \path (0) edge[draw=red, text=red]  node {$\bigstar$} (2);
      \path (1) edge[draw=red, text=red, above right]  node {$\bigstar$} (2);
      \path (0) edge[draw=blue, text=blue, below left]  node {$I \to O$} (3);
      \path (1) edge[draw=blue, text=blue]  node {$I \to O$} (3);
      \path (2) edge[draw=red, text=red, loop above]  node {$\bigstar$} (2);
    \end{tikzpicture}
  \end{minipage}
  \caption{Checkable self-closing door construction (left) and gadget prior to post-selection (right).
  The entrance is $A$, exit is $B$, and keyhole is $C$.
  The door starts open, becomes closed when the agent travels from $A \to B$, and opens back up when the agent enters and exits $C$.
  The checking path begins by closing off $A$, $B$, and $C$.
  After that, the agent checks that the upper 1-toggle has the correct parity (indicated by the dotted blue arrow), and then proceeds through the checking transitions of each checkable sub-gadget.
  $\bigstar$ is used here to indicate any non-trivial transition between $\{A, B, C\}$ and $\{A, B, C\}$, with the exception of those which are intended.
  All of these breaking transitions permanently prevent the agent from checking the gadget.}
  \label{fig:scd}
\end{figure}

Using the previously constructed gadgets, we build a variant of 
\textit{Pushing Blocks via Checkable Gadgets}'s~\cite{ani22pushing} two diodes, one 1-toggle, and a precursor blueprint for a self-closing door.
Here, for the sake of formal verification, we include the explicit check-path.
This construction and gadget is depicted in Figure~\ref{fig:scd} and is formally verified to exactly
match the semantics of a self-closing door.

\noindent By the construction and verification of this sequence of gadgets, Push-1 is PSPACE-complete. $\square$

 \section{Discussion}

Building on prior techniques, this work proves that Push-1 is PSPACE-complete, answering one of the two remaining questions about the Push family of problems.
This leaves the complexity of Push-\textasteriskcentered\ an open question to be tackled by future work.

Outside of the Push family, this paper's proposed extension to the classical motion-planning-through-gadgets framework and its system for formally verifying constructions in this framework will hopefully prove useful in resolving other open problems related to the complexity of reachability.\\

\noindent The code for \sysname is publicly available at \href{https://github.com/mathmasterzach/gadgeteer}{github.com/mathmasterzach/gadgeteer}.

\paragraph{Addendum}
Concurrent to the initial release of this preprint, an independent proof of the PSPACE-completeness of Push-1 reachability appeared~\cite{mit2025push}.
While the two proofs follow substantially different reductions and make interesting independent contributions, they have converged on similar techniques for Push-1 constructions.
We note that both use check-paths which eject blocks into the bodies of constructions to ensure they have not been broken.
Additionally, both construct a hallway cutoff and a closing crossover (in their work called a ``single-use closing'' and a ``strong closing crossover'' respectively) using similar systems of gadgets.
All of the constructions in this alternate proof are amenable to verification in \sysname.

\subsubsection*{Acknowledgements}

Various members of the \href{https://pathology.thinky.gg}{Pathology} community (a community centered around a game with similar rules to Push-1 located at the URL \href{https://pathology.thinky.gg}{pathology.thinky.gg}) aided in the low-level development of constructions, in the testing of \sysname, and in the presentation of the results. This paper was improved immensely by their contributions, in particular by the detailed suggestions and feedback of David Spencer, George Spahn, 
Michael Gottlieb, and Spencer Spenst.
Figures were produced with input from \href{https://github.com/edemaine/svgtiler}{SVG-Tiler}.

\bibliographystyle{plain}
\bibliography{references}

\end{document}